\documentclass[aps,prl,showpacs,twocolumn]{revtex4}
\usepackage{epsfig}
\usepackage{color}

\begin{document}

\title{Electrostatics of straight and bent nanotubes}

\author{E. G. Mishchenko}
\affiliation{Department of Physics, University of Utah, Salt Lake
City, UT 84112}

\author{M. E. Raikh}
\affiliation{Department of Physics, University of Utah, Salt Lake
City, UT 84112}

\begin{abstract}
Response of a single-walled carbon nanotube to external electric
field, $F$, is calculated analytically within the classical
electrostatics. Field-induced charge density distribution is
approximately linear along the axis of a {\em metallic}  nanotube
and depends rather weakly,
 as $\ln (h/r)$, on the nanotube length, $h$,
 (here $r$ is the
nanotube radius). In a semiconducting nanotube with a gap, $E_g$,
charge separation occurs as $F$ exceeds the threshold value
$F_{th}=E_g/eh$. For $F>F_{th}$, positively and negatively charged
regions at the ends of nanotube are separated by a {\em neutral
strip} in the middle. Properties of this neutral strip, length and
induced charge distribution near the ends, are studied in detail. We
also consider a bent nanotube and demonstrate that the number of
neutral strips can be one or two depending on the direction of $F$.
\end{abstract}


\maketitle

\section{Introduction} There are two types of devices based on carbon
nanotubes (NTs) that are currently intensely investigated. These are
field-effect transistors and field emitters for flat panel displays
and x-ray sources. Although both types of devices were reported
several years ago \cite{tans,Science0} and significant improvement
in their characterisitcs had been recently achieved~ (see {\em
e.g.}, Refs.~\cite{Javey,mceuen05,dai05} and
Refs.~\cite{Science1,mauger05}),
%still the understanding of
%their characterisitcs are steadily improving
%\cite{Javey,mceuen05,Science1,mauger05},
 the understanding of
performance of these devices is far from complete. The prime reason
for this is  1D-like geometry of NT-based transistors and field
emitters. Due to this geometry, their electrostatics is
qualitatively different from that in respective well-understood 2D
counterparts.
%that electrostatics
%of NT-based transistors and field emitters, which have 1D-like
%geometry, is qualitatively different from their
%well-understood 2D counterparts.

Traditionally,  electrostatics for particular geometry  of NT
transistor \cite{datta,ratner} or field emitter \cite{mayer,buldum}
is studied theoretically by employing a certain version of
first-principle calculations. Notable exceptions are
Refs.~\cite{odintsov,odintsov1}, in which classical electrostatics
was used to describe the potential profile near the contact between
metallic and semiconducting NTs as well as the contact between NT
and a metal. Applicability of this description of the contact
phenomena in NTs was later questioned in Ref.~\cite{tersoff}.

In general, it is not obvious whether or not, in terms of
electrostatics, a NT of a small radius, $r$, can be modeled by an
infinitely thin sheet of electron gas wrapped into a cylinder. In
other words, whether or not the potential distribution can be
adequately described by the Poisson equation with boundary
conditions imposed at distance, $r$, from the NT axis. Positive
answer to this question was recently given in Ref.~\cite{classical},
where the density-functional calculations of the extra charge
distribution along the NT were shown to be in quantitative agreement
with classical electrostatics analysis.

In this situation, it is instructive to consider a model problem of
the classical electrostatics of a NT, which allows for an {\em
analytical} solution. Qualitative features of this solution might
then yield a valuable insight into electrostatics of realistic
devices. Such a problem is studied in the present paper. Namely, we
consider a NT in external electric field, $F$, parallel to the NT
axis. Separation of variables in the Poisson equation in this
geometry is impossible. Still, as we demonstrate below, presence of
a small parameter, $r/h$, where $h$ is the NT length, allows one to
obtain the {\em asymptotically exact} distribution of potential.
%For such a geometry, the asymptotically exact
%distribution of potential can be found by virtue of a small
%parameter, $h/r$, where $h$ is the NT length.
We show that for {\em metallic} NT the density of induced charge
changes linearly with distance from the NT center. For {\em
semiconducting} NT with a gap, $E_g$, charge separation,  which
occurs as $F$ exceeds the threshold value $E_g/eh$, results in
formation of a {\em neutral strip} with a width $\approx E_g/eF$ in
the center of the NT. We find the profile of the charge density
growth from the edges of the strip towards the NT ends. Finally, we
use the developed approach to describe quantitatively the
electrostatics of bent or wiggly NTs in external field, pertinent to
recent electroabsorption measurements \cite{Vardeny}, and
demonstrate that wiggling results in multiple alternating positively
and negatively charged regions separated by neutral strips.

\section{Basic equation}

Denote with $\rho(z)$ the {\em linear} density of charge, induced by
external field on the NT surface. Then the local value of the Fermi
momentum is given by $p_{F}(z)=\pi\hbar\vert \rho(z)\vert/2eN$,
where $N$ accounts for the spin and band degeneracy ($N=2$ or $4$,
and is determined by the NT chirality). The local chemical
potential, $\mu(z)$, is related to $p_{F}(z)$ via the NT energy
spectrum
\begin{equation}
\label{spectrum} \mu(z)=\text{sign}(z)
\sqrt{{E_g^2}/{4}+v_0^2p_F^2(z)} ,
%\left[\frac{E_g2}{4}+v_0^2p_F2(z)\right]^{1/2} ,
\end{equation}
where $v_0\approx 8\times 10^7$ cm/s is the electron velocity in
graphene. Second relation between $\mu(z)$ and $\rho(z)$ expresses
the fact that the electrochemical potential remains constant along
the nanotube, i.e. $\mu(z)+e\varphi(z)=0$, where $\varphi(z)$ is the
electrostatic potential
\begin{equation}
\label{potential}
\varphi(z)=-Fz+\frac{1}{e}\int_{-h/2}^{h/2}dz^{\prime}\rho(z^{\prime})
\Phi(z-z^{\prime}),
\end{equation}
which is created by external field and by induced charges. The
kernel, $\Phi(z-z^{\prime})$, in Eq.~(\ref{potential}) takes a
simple form in the case of  isolated NT, lying on substrate with
dielectric constant, $\varepsilon$
\begin{equation}
\label{interaction} \Phi(x)= \frac{e}{\pi\varepsilon^{\ast}}
\int_0^{\pi}\!\frac{d\alpha}
{\left[x^2+4r^2\sin^2(\alpha/2)\right]^{1/2}}~~,
\end{equation}
where $\varepsilon^{\ast}=\left(\varepsilon +1\right)/2$.
%$\varepsilon$ being the dielectric constant of the substrate.
%In the presence of a gate at a distance, $D$, from the NT,
%the image chages should be taken into account, so that
%$\Phi(x)=
%\Phi_0(x)-\Phi_0\left(\sqrt{x2+4D2}\right)$.
With the help of Eqs.~(\ref{spectrum}) and (\ref{potential}), the
condition of constant electrochemical potential can now be presented
as a  {\em closed} integral equation for $\rho(z)$
\begin{equation}
\label{closed} eFz=\sqrt{\frac{E_g^2}{4}+ \left[\frac{\pi\hbar
v_0\rho(z)}{2eN}\right]^2}\!\!+
\!\int_{-h/2}^{h/2}\!\!\!\!dz^{\prime}\rho(z^{\prime})
\Phi(z-z^{\prime}),
\end{equation}
where we assumed $z$ to be positive. Eq.~(\ref{closed}) should be
complemented by the obvious condition that $\rho(z)$ is {\em odd}.

\section{Asymptotic solution of Eq.~(\ref{closed})}

In order to make use of the small parameter $r/h$, we rewrite the
integral on the rhs of Eq.~(\ref{closed}) as follows
\begin{equation}
\label{byparts}
\int_{-h/2}^{h/2}\!\!\!\!dz^{\prime}\rho(z^{\prime})\Phi(z-z^{\prime})=
\int_{0}^{h/2}\!\!\!\!dz^{\prime}\frac{\partial\rho(z^{\prime})}
{\partial z^{\prime}}K(z,z'),
\end{equation}
where the function $K(z,z')$ is defined as
\begin{equation}
\label{defined}
K(z,z')=\int_{z^{\prime}}^{h/2}dz_1\Bigl[\Phi(z-z_1)-
\Phi(z+z_1)\Bigr].
\end{equation}
Our most important observation is that, in the limit $h \gg r$, the
function  $K(z,z')$ can be replaced by
$\left(2e/\varepsilon^{\ast}\right)\ln(h/4r)\Theta(z-z^{\prime})$,
where $\Theta(x)$ is the step-function. Possibility of such a
replacement is illustrated in Fig.~1.
\begin{figure}[h]
\resizebox{.40\textwidth}{!}{\includegraphics{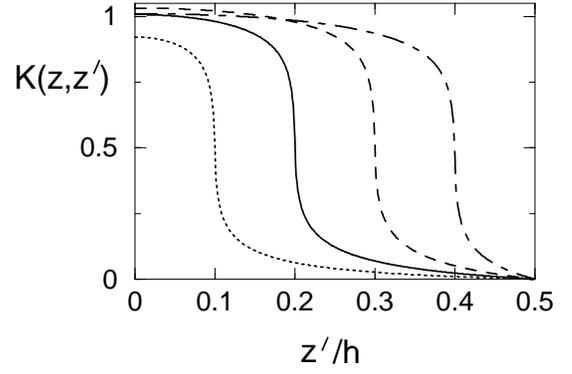}}
\caption{Plot of the function $K(z,z')$ [scaled with the factor
$2e\ln(h/4r)/\varepsilon^{\ast}]$ for $h/r=10^3$ and different
values of $z$: $z=0.1h$ (dotted line), $z=0.2h$ (solid line),
$z=0.3h$ (dashed line), $z=0.4h$ (dash-dotted line).}
\end{figure}
A simple from of $K(z,z')$ allows for a drastic simplification of
Eq.~(\ref{closed}), which transforms from intergral equation to a
simple algebraic (quadratic)  equation. In particular case, $E_g=0$
(metallic NT), we obtain the following result for the induced charge
distribution
\begin{equation}
\label{simplest}
\rho(z)\approx %sign(z)
\frac{\varepsilon^{\ast}g Fz}{1+2g\ln(h/r)},
\end{equation}
where we have introduced a dimensionless interaction parameter
$g=2Ne^2/\pi\varepsilon^{\ast}\hbar v_0$. The above result for
$\rho(z)$ has a logarithmic accuracy, in the sense, that  numerical
factor in the argument of logarithm is not specified. In particular,
the {\em height} of the step-function in Fig.~1 contains $1/4$ under
the logarithm. Another contribution to the argument of logarithm
comes from the {\em smearing} of the step-function in Fig.~1. This
smearing gives rise to the relative correction $-2z\partial \ln
\rho(z)/\partial z$ to $\ln (h/4r)$;
%so that
in particular case of Eq.~(\ref{simplest})
%the factor under the logarithm is $1/4e2$.
this correction is equal to $-2$. Overall, the condition of
applicability of Eq.~(\ref{simplest}) is $\ln(h/r) \gg 1$, which is
met in  most of the realistic situations. It follows from
Eq.~(\ref{simplest}) that  the polarizability, $\chi$, of the  NT,
defined as $P(F)\!=\!\int_{-h/2}^{h/2}dz z\rho(z)=\chi F$, has the
form
\begin{equation}
\label{polarizability} \chi=\frac{\varepsilon^{\ast}g
h^3}{12\Bigl[1+2g\ln(h/r)\Bigr]}.
\end{equation}
It is clear from Eq.~(\ref{polarizability})  that the product
$g\ln(h/r)\approx (1.74N/\varepsilon^{\ast})\ln(h/r)$ is a
quantitative measure of the ``metallicity''of the NT. In the limit
of a ``long'' NT, when the product $g\ln(h/r)$ is large, we have
$\chi =\varepsilon^{\ast}h^3/24\ln(h/r)$, which coincides with the
textbook expression \cite{Landau} for polarizability of a perfectly
conducting ellipsoid with axes $r$ and $h\gg r$. In this limit, with
external field  parallel to the NT axis, the {\em resulting} field
at the NT surface is {\em normal} to this surface. The opposite
limit, $2g \ln(h/r) \ll 1$, of a ``short'' nanotube, when external
field is altered weakly by the induced charges, cannot  be achieved
even for high dielectric constant of the substrate, {\em e.g.},
$\varepsilon^{\ast}\approx 6$ for Si.

Consider now a semiconducting (or strained metallic \cite{strain})
NT with finite $E_g$. It is seen from Eq.~(\ref{closed}) that charge
separation occurs only when the external field is strong enough,
namely, $F> F_{th}=E_g/eh$. It also follows from Eq.~(\ref{closed})
that, as $F$ increases, electrons and holes emerge at the NT {\em
ends}, while the strip $\vert z\vert < E_g/2eF$ in the center of NT
remains {\em neutral}. The behavior of $\rho(z)$ outside the strip
is given by
\begin{eqnarray}
\label{long}
\rho(z)\!=\!\left(\frac{\varepsilon^{\ast}E_g}{2e}\right)\!\!
\Biggl[g^2\ln^2(h/r)-\frac{1}{4}\Biggr]^{-1}\times
\nonumber\\
\Biggl[\!g^2\ln(h/r)\!\!\left(\frac{eFz}{E_g}\right)\!-\!\frac{g}{2}\sqrt{\left(\frac{eFz}{E_g}\right)^2\!\!+\!\!
g^2\ln^2(h/r)-\frac{1}{4}}\Biggr],
\end{eqnarray}
and is illustrated in Fig.~2. From the edge of the neutral strip to
the ``bulk'' of NT the  slope of $\rho(z)$ decreases by a factor
$2g\ln(h/r)/[2g\ln(h/r)+1]$. Using Eq.~(\ref{long}), one can
calculate the induced dipole moment, $P(F)$, of semiconducting NT.
Obviously, for $F\gg F_{th}$ it is the same $\chi F$ as for metallic
NT. In the vicinity of the threshold, $(F-F_{th})\ll F_{th}$, the
induced charge density not only occupies small region near the tips,
but is also small in magnitude. Therefore, $P(F)$ is quadratic in
$(F-F_{th})$ near threshold, namely
\begin{equation}
\label{threshold}
P(F)=\frac{\varepsilon^{\ast}h^3F_{th}}{4\ln(h/r)}\left(\frac{F}{F_{th}}-1\right)^2.
\end{equation}
\begin{figure}[h]
\resizebox{.28\textwidth}{!}{\includegraphics{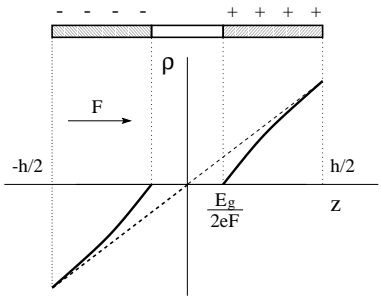}}
\caption{Charge density distribution induced by external field, $F$,
along the metallic (dashed line) and semiconducting (solid line)
NT.}
\end{figure}\\

\section{Small $E_g$; Fine structure of the neutral strip}

The boundaries of the neutral strip, $z_t=\pm E_g/2eF$, were found
from Eq.~(\ref{closed}) within the ``local'' approximation. There
are two sources of corrections to this result: classical and
quantum. Classical correction originates from the fact that for
small $E_g$ the neutral strip is surrounded by long charged regions
with opposite signs of charge. Setting $z=z_t$ in
Eq.~(\ref{closed}), and substituting into the rhs the zero-order
result (\ref{long}) for $\rho(z)$, we obtain the following  modified
equation for $z_t$
\begin{equation}
\label{modified}
z_t=\frac{E_g}{2eF}+\frac{2g\ln(h/z_t)}{1+2g\ln(h/r)}z_t.
\end{equation}
The second  term in the rhs of Eq.~(\ref{modified}) reflects the
fact that potentials, created by the left and right charged
neighbors of the neutral strip, do not compensate each other
completely. As follows from Eq.~(\ref{modified}), the relative
classical correction
 to the position of the boundary, $z_t$, is
 $\delta z^{cl}_t/z_t\approx 2\ln{(h/z_t)}/\ln{(h/r)}$. This correction
is small provided  that $eFr \ll E_g$. The latter condition also
insures that the underlying energy spectrum of the NT is not
affected by $F$, as was assumed in derivation of our basic equation
(\ref{closed}). Quantum correction to $z_t$ comes from the
penetration of electronic wave functions into the classically
forbidden region inside the neutral strip. Using Zener's formula for
tunneling exponent in the Dirac spectrum (\ref{spectrum}), the
relative quantum correction can be presented in the form, $\delta
z^{q}_t/z_t\sim (l_B/gz_t)^{2/3}$, where $l_B=e^2/E_g\varepsilon^*$
has the meaning of the exciton Bohr radius. Thus, the condition of
smallness of the quantum correction to $z_t$ coincides with the
condition that $F$ is weaker than the ionization threshold for an
exciton.

\section{$F=0$; Charged metallic NT}

Experimental situations in which electrostatics of a charged NT is
important are listed in Ref.~\cite{classical}. Classical
electrostatic analysis of a charged NT was carried out in this paper
only for a short NT (with $h/r \approx 4$). Below we find the
distribution of charge analytically in the limit $h \gg r$. For a
charged metallic NT the condition of a constant electrochemical
potential takes the form
\begin{equation}
\label{charged} \mu= \left(\frac{\pi\hbar v_0}{2eN}\right)
\rho(z)+\int_{-h/2}^{h/2}
dz^{\prime}\rho(z^{\prime})\Phi(z-z^{\prime}).
\end{equation}
Now $\rho(z)$ is even. To employ the above ansatz, we take
derivative from the both sides of Eq.~(\ref{charged}) and perform
integration by parts in the rhs. This yields
\begin{equation}
\label{by} \rho(h/2) \left[\Phi_{\mbox{\tiny$-$}} -
\Phi_{\mbox{\tiny$+$}}\right]=
%\frac{\pi\hbar v_0}{2eN}
\frac{e R(z)}{g\varepsilon^*}
+\int_{-h/2}^{h/2}\!\!\!\!\!dz^{\prime}
R(z^{\prime})\Phi(z-z^{\prime}),
\end{equation}
 where $R(z)=\partial\rho(z)/\partial z$ is the {\em odd} function
of $z$, and a short-hand notation $\Phi_\pm(z)=\Phi(\frac{h}{2}\pm
z)$ is introduced. Now Eq.~(\ref{by}) has the form similar to
Eq.~(\ref{closed}). It should be complemented by the condition
$2\int_{0}^{h/2}dz \rho(z)=Q$, where $Q$ is the total charge on the
NT. Using the similarity between Eqs.~(\ref{by}) and (\ref{closed}),
we readily obtain
\begin{equation}
\label{Q} \rho(z)=\frac{Q}{h}\Biggl[1+\frac{g}{1+g\ln(h/r)}
\ln\left(\frac{h^2}{h^2-4z^2}\right)\Biggr].
\end{equation}
Hence,
%It is seen from Eq.\ (\ref{Q}) that
$\rho(z)$ is essentially constant along the NT \cite{french}, and
raises sharply only near the tips. This behavior compares favorably
with numerical results \cite{classical}. Logarithmic divergence in
Eq.~(\ref{Q}) is terminated in the vicinity of the tip $ h/2-z
\lesssim r$, so that the net growth of $\rho(z)$ is given by
$\rho(h/2)/\rho(0)=[1+2g\ln(h/r)]/[1+g\ln(h/r)]$.

In optical experiments on separated NTs the tubes usually have
wiggly shapes. Then their response to the external field can be
quite peculiar, with numerous alternating positively and negatively
charged regions separated by neutral strips (see Fig.~3a). An
insight into electrostatics of a wiggly NT can be drawn from a model
example that allows for exact solution.

\section{Bent NT}

 We consider a NT in the form of a semicircle
of a radius, $R$. In two limiting cases, the electric field is
either pointed along the diameter connecting the NT tips  (parallel
geometry) or perpendicular to this diameter (perpendicular geometry,
see Fig.~3).
\begin{figure}[h]
\resizebox{.42\textwidth}{!}{\includegraphics{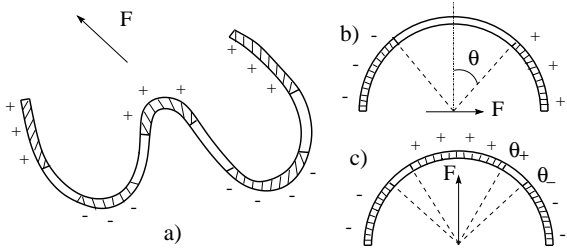}}
\caption{Schematic illustration of charge separation in
semiconducting NT
 in external field: (a) NT of a wiggly shape; (b) and (c) NT of a semicircle shape with $F$
parallel and perpendicular to the diameter, respectively.}
\end{figure}
Both geometries are described by Eq.~(\ref{closed}) written in polar
coordinates
\begin{eqnarray}
\label{closed_bent_perp} && eF_t(\theta)R -{\cal
C}=\text{sign}(\rho)\sqrt{{E_g^2}/{4}+ [e \rho(\theta)/g\varepsilon^*]^2}\!\!\nonumber\\
&&+\frac{eR}{\varepsilon^*} \!\int_{-\pi/2}^{\pi/2}
\!\!\!d\theta^{\prime}\rho(\theta^{\prime}) ~ \Phi\left(R\sin{
\frac{ |\theta-\theta^\prime|}{ 2}}\right),
\end{eqnarray}
where $F_t(\theta)$ is the tangent component of $F$. The constant,
${\cal C}$, which determines the electrochemical potential of NT,
must be found from the condition of the net NT neutrality,
$\int_0^\pi d\theta \rho(\theta)=0$. We emphasize that in {\it both}
the parallel and perpendicular geometries only the tangent component
of electric field is responsible for the charge separation. The
normal component field has no effect on the charge separation if the
condition $E_g \gg eFr$ is satisfied. The latter condition means
that field-induced mixing of transverse subbands is negligible.

 \noindent(i) In  parallel geometry, we have
$F_t=F\sin{\theta}$. Thus, $\rho(\theta)$ is {\em odd}, so that
${\cal C}=0$. Same ansatz as for a straight NT, yields the following
solution of Eq.~(\ref{closed_bent_perp}) for $E_g=0$
$\rho(\theta)=\varepsilon^{\ast}gFR\sin{\theta}/ [1+2g\ln(R/r)]$. It
is also easy to see from Eq.\ (\ref{closed_bent_perp}) that, for
finite $E_g$, the neutral strip occupies the segment $|\theta|<
\arcsin{(2eFR/E_g)}$ of the NT near its top (Fig.~3b).

\noindent(ii) In perpendicular geometry,
$F_t(\theta)=F\cos{\theta}$,  is the {\em even} function of
$\theta$. As a result, $\rho(\theta)$ differs qualitatively from the
case of a straight NT. The form of $\rho(\theta)$ for this geometry
can be found from the same ansatz Eq.~(\ref{by}) that was used for
the straight charged NT. In particular, for $E_g=0$ we obtain
\begin{equation}
\label{perp}
\rho(\theta)=\varepsilon^*FR~\frac{g\left(\cos{\theta}-2/\pi\right)}{1+2g\ln{(R/r)}}.
\end{equation}
The {\em two} points  at which $\rho(\theta)$ changes the sign are
therefore located at $\theta=\pm\theta_0=\pm 50.5^{\circ}$. For a
finite $E_g$ two neutral strips are formed around
$\theta=\pm\theta_0$. Their boundaries, $\pm\theta_{+}$ and
$\pm\theta_{-}$, (see Fig.~3c) are determined by the conditions
\begin{equation}
\label{pm} \cos\theta_{+}= \frac{2{\cal
C}+E_g}{2eFR},~~~\cos\theta_{-}= \frac{2{\cal C}-E_g}{2eFR}.
\end{equation}
When the gap is small, $E_g\ll eFR$, the centers of the strips are
still located near $\pm \theta_0$, while the strip width,
$\delta\theta=(\theta_{\mbox{\tiny$-$}}-\theta_{\mbox{\tiny$+$}})$,
can be found directly from Eq.~(\ref{pm}), namely $\delta\theta =
E_g/eFR\sin\theta_0\approx 1.3E_g/eFR$.

For small ratio $E_g/2eFR$ it is also easy to trace the crossover
between the parallel and perpendicular geometries as the field is
rotated. For rotation angle, $\beta$, and for $E_g=0$, a
straightforward generalization of Eq.~(\ref{perp}) yields the
following angular dependence of the charge density $\rho(\theta)
\propto \left[\cos(\beta-\theta)-\frac{2}{\pi}\cos\beta\right]$.
From this dependence we conclude that, as $\beta$ decreases from
$90^{\circ}$ (parallel geometry), the  narrow neutral strip at the
top of semicircle moves to the left.  At critical $\beta_c=\arctan
(2/\pi)\approx 32.5^{\circ}$, when the neutral strip is located
around $\theta\approx 25^{\circ}$, the second neutral strip emerges
at the right end of the NT. As $\beta$ decreases further, both
neutral strips move to the left and assume their ``perpendicular''
positions $\theta = \pm 50.5^{\circ}$.\\

\section{Experimental implications}

 In
Ref.~\onlinecite{distribution} the potential distribution along the
NT was measured using the atomic force microscope. Experimentally
measured profile of the voltage drop in the structures with a small
contact resistance is consistent with existence of a neutral strip
near the NT center. Note, that the substrate in
Ref.~\onlinecite{distribution} was thick: $D=200$ or $D=500$nm, but
still thinner than the NT length, $h=1200$nm. Taking into account
the presence of a gate at distance $D \lesssim h$ from the NT
amounts to replacement \cite{rotkin,sapmaz,latessa,anantram06} of
$\ln (h/r)$ by $\ln (D/r)$ in all
the above formulas.\\
 Our results have direct relevance to the electro-optics of NTs.
Measurements of electroabsorption in single-walled carbon NTs were
recently reported in Ref.~\onlinecite{Vardeny}. It might seem that
with photon energy $\sim 1$eV much bigger than $E_g$ the large-scale
nanotube geometry is not important.  This, however, is not the case.
The reason is that the dipole moment of the many-body optical
transition \cite{Kane,exciton} is directed {\em along} the tube
\cite{exciton,Liang}. Even if external field is parallel to the tube
axis, the {\em resulting} field is almost perpendicular to the
nanotube surface, and thus its  effect on the optical transitions is
suppressed. This strong suppression by a factor $[1+ 2g\ln(h/r)]$
must be taken into account when the oscillator strength is extracted
from electroabsorption \cite{Vardeny}. On the other hand, within the
neutral strip, the acting field is equal to the applied field.
However, the relative contribution,  $E_g/eFh$, of neutral strips to
the absorption is small.\\
As a final remark, note that dimensionless parameter, $g$, which
governs the screening properties of the  NT, has a transparent
meaning. For 2D electron gas with a density of states, $\nu$, the
linear screening length is equal to $l=\varepsilon^{\ast}/2\pi\nu
e^2$. If the gas is wrapped into a cylinder of a radius, $r$, then
the degree of penetration of external field inside the cylinder is
determined by the ratio $r/l$. Up to a numerical coefficient, this
ratio is nothing but the parameter $g$.

\section{Acknowledgements}

 This work was supported by NSF under
Grant No. DMR-0503172 and by the Petroleum Research Fund under Grant
No. 43966-AC10. Useful discussions with Z. V. Vardeny on optics of
NTs are acknowledged.

\end{document}